\documentclass[twocolumn,superscriptaddress,showpacs,preprintnumbers,amsmath,nofootinbib,amssymb]{revtex4}

\usepackage{epsfig}
\usepackage{graphicx}
\usepackage{dcolumn}
\usepackage{bm}
\usepackage{color}

\def\thebiblio#1{
\begin{center}\bf \large References
\end{center}
\list
{[\arabic{enumi}]}{\settowidth\labelwidth{#1.}\leftmargin\labelwidth
 \advance\leftmargin\labelsep
 \usecounter{enumi}}
 \def\newblock{\hskip .11em plus .33em minus -.07em}
 \sloppy
 \sfcode`\.=1000\relax}

\renewcommand{\[}{\begin{equation}}
\renewcommand{\]}{\end{equation}}

\begin{document}



\title{Warm hilltop inflation}


\author{Juan~Carlos~Bueno~S\'anchez}%
\email{j.buenosanchez@lancaster.ac.uk}
\affiliation{%
Physics Department, Lancaster University, Lancaster LA1 4YB, UK}%
\author{Mar~Bastero-Gil}%
\email{mbg@ugr.es}
\affiliation{%
Departamento de F\'{\i}sica Te\'orica y del Cosmos,
Universidad de Granada, Granada-18071, Spain}%
\author{Arjun~Berera}%
\email{ab@ph.ed.ac.uk}
\affiliation{%
School of Physics, University of Edinburgh, Edinburgh EH9 3JZ, UK}%
\author{Konstantinos~Dimopoulos}%
\email{k.dimopoulos1@lancaster.ac.uk}
\affiliation{%
Physics Department, Lancaster University, Lancaster LA1 4YB, UK}%

\date{\today}



\begin{abstract}
We study the low-temperature limit of warm inflation in a hilltop
model. This limit remains valid up to the end of inflation,
allowing an analytic description of the entire inflationary stage.
In the weak dissipative regime, if the kinetic density of the
inflaton dominates after inflation, low scale inflation is
attained with Hubble scale as low as 1~GeV. In the strong
dissipative regime, the model satisfies the observational
requirements for the spectral index with a mild tuning of the
model parameters, while also overcoming the $\eta$-problem of
inflation. However, there is some danger of gravitino
overproduction unless the particle content of the theory is large.
\end{abstract}


\maketitle



\section{Introduction}

The latest high-precision observations strongly suggest that the
Universe, during its early history, went through a period of
almost exponential expansion called inflation. Inflation not only
accounts for the horizon and flatness problems of the Big Bang
cosmology, but also provides the primordial curvature perturbation
which seeds the formation of structure and is observed through the
anisotropy of the cosmic microwave background (CMB) radiation. It
is a major success of inflation that the CMB observations confirm
the existence of superhorizon curvature perturbations, which are
predominantly adiabatic and Gaussian with an almost
scale-invariant spectrum.

The standard paradigm for realising inflation in the context of particle
theory employs a scalar field, called inflaton, whose potential density
dominates the Universe \cite{book}.
The dynamics of this scalar field is governed by its
scalar potential, the form of which determines the dynamics of inflation.
In order to produce an almost scale-invariant spectrum of curvature
perturbations the potential density has to remain almost constant, which means
that the scalar field should vary extremely slowly during inflation. This, in
turn, is possible only if the potential is sufficiently flat. Hence, the
inflaton field is usually assumed to be a flat direction in field space.
In order to avoid lifting the flatness of the potential due to radiative
corrections the inflaton is typically assumed to be a gauge singlet
(see however \cite{mssm}) with suppressed interactions to other fields.

One of the most important discoveries of the CMB observations is
the fact that the spectrum of the curvature perturbation deviates
significantly from the exact scale invariance of the
Harrison-Zel'dovich, ``vanilla'' case. Instead, the spectrum
appears to be red, with spectral index:
\mbox{$n=0.948^{+0.016}_{-0.015}$} within the 1-$\sigma$ window
\cite{Spergel:2006hy} (when tensors and running of $n$ are
negligible). This result reveals that the physics of inflation is
non-trivial and enables model-builders to discriminate between
models. Indeed, it seems for example, that the simplest form of
hybrid inflation \cite{hybrid}, which produces a blue spectrum, is
already excluded (However, supersymmetric hybrid inflation models
\cite{Copeland:1994vg,shybrid} produce a red spectrum and can
still be in agreement with observations \cite{Alabidi:2005qi}).
Also, many large-field models, such as quartic (or higher-order)
chaotic inflation are also incompatible with observations
\cite{Alabidi:2005qi}. In terms of simple, single-field models,
one class that appears to be generically in good agreement with
CMB observations is the so-called small-field models, the
prototype of which is `new inflation' \cite{new}. In these models,
the inflaton is rolling off the top of a potential hill. Hence,
the realisation of this scenario in terms of particle theory
corresponds to a number of inflation models dubbed `hilltop
inflation' \cite{hilltop1,hilltop2}. In general, hilltop inflation
produces a red spectrum of curvature perturbations and negligible
tensors (tiny tensor fraction $r_t\lesssim 0.002$). The variation
of the inflaton field is much smaller than the Planck scale, which
means that the scalar potential is much more understandable in
terms of effective field theory \cite{hilltop1}.

However, hilltop models have their own problems. Firstly, if they
are to generate the curvature perturbation, then the curvature of
the top of the potential hill has to be fine-tuned, otherwise the
spectrum becomes too much tilted. However, in the context of
supergravity and superstring theories, such fine-tuning is
unnatural. This is the so-called $\eta$-problem plaguing most
models of inflation. Slow-roll inflation requires \mbox{$|\eta|\ll
1$} in order to produce an approximately scale-invariant spectrum,
while supergravity corrections to the scalar potential (barring
accidental cancellations) suggest \mbox{$|\eta|\sim 1$}
\cite{Dine:1995uk}. Such a large $|\eta|$ would result in
fast-roll inflation \cite{FR}, which works only if the curvature
perturbations are generated by a field other than the inflaton
(e.g. a curvaton field \cite{curv}). In this paper, we show that
dissipative effects may eliminate the $\eta$-problem of hilltop
inflation.

Another problem of hilltop models has to do with the necessary initial
conditions. The inflaton field must find itself near the top of the potential
hill in order for inflation to occur. Some authors argue that the initial
condition problem is evaded by considering the possibility of eternal
inflation \cite{hilltop1}.
It is true that near the top of the potential hill there is a
region of quantum diffusion which would result in eternal inflation were the
field placed there originally. Eternal inflation then guarantees that the
regions of the multiverse which have undergone inflation occupy much more
volume than the ones that have not, so it appears much more likely for an
observer to find oneself in one of those. Hence, if we envisage random initial
conditions, the chances are that, at late times, the most probable locations
would be the ones which did inflate and hence the ones where the initial
conditions corresponded to the top of the hill. The weak point of this argument
is that, in this setup, by far the largest probability at late times is to find
oneself in a region that is still inflating and, therefore, occupies much more
volume than the regions which stopped inflating 14 billion years ago. One then
needs to employ anthropic arguments to place our location in the multiverse in
the non-inflating region. So the eternal inflation explanation of the initial
conditions for hilltop inflation relies on anthropic reasoning, which many
authors do not find convincing.

Traditionally, the initial conditions for new inflation have been
addressed by assuming that the top of the potential hill
corresponds to a fixed point of the symmetries of the theory. In
fact, in string moduli space, we can envisage that the fixed point
is a point of enhanced symmetry. In this case, even if the field
is originally rolling past this point it is highly probable that
the enhanced interactions will temporarily trap it there
\cite{trap} accounting thereby for the appropriate initial
conditions. In this paper we explore further the implications of
such interactions to the dynamics of hilltop inflation. If the
inflaton's couplings to other degrees of freedom are not
negligible then we may expect the inflaton to perturbatively
dissipate some of its energy into other fields, generating thereby
a thermal bath during inflation. Hence, we explore whether
dissipation effects may give rise to a warm inflation scenario
\cite{warm}.

There are several attractive model building features found in warm
inflation scenarios. For one, due to dissipative effects, the
inflaton mass during inflation can be much bigger than the Hubble
scale \cite{Berera:1999ws, Berera:2004vm}, thus completely
avoiding the $\eta$-problem
\cite{Copeland:1994vg,Dine:1995uk,Gaillard:1995az,
randall},
which is a generic problem in standard cold inflation
scenarios in supergravity theories. Another attractive model building
feature is for monomial potentials, inflation occurs with the
inflaton amplitude below the Planck scale \cite{Berera:1999ws,
Berera:2004vm}. In contrast, for monomial potentials in the cold
inflation case, usually called chaotic inflation scenarios
\cite{ci}, the inflaton amplitude during inflation is larger than
the Planck scale.  This is a problem for model building, since in
this case the infinite number of nonrenormalizable operator
corrections, $\sim \sum_{n=1}^{\infty} g_n \phi^{4+n}
(\phi/m_{P})^n$ would become important and so have to be retained
\cite{randall}.

We study a wide range of dissipative effects from weak to strong.
In the entire range attractors are found, where during inflation
there exists a subdominant
thermal bath, with roughly constant temperature maintained by the perpetual
perturbative decay of the inflaton. The temperature of this thermal bath is
higher than the Hawking temperature corresponding to the (quasi)de Sitter
expansion. As a result the curvature perturbation is due to the thermal
fluctuations of the inflaton field which dominate over the quantum ones.
In the strong dissipation regime the density of the thermal bath eventually
takes over the potential density of the inflaton thereby reheating the
Universe. In contrast, if dissipation is weak, inflation has to be terminated
dynamically, by assuming that the potential steepens further when the field
is far enough from the origin, in accordance to hilltop inflation models.
To model the cosmology after the end of inflation, in this case, we assume
that the inflaton becomes kinetically dominated and drives a brief period of
kination \cite{kin} until its density is taken over by the thermal bath, which
reheats the Universe. This is a scenario which naturally arises in models where
the inflaton is characterised by a runaway potential (e.g. a string modulus)
and has been employed in models of quintessential inflation \cite{eta,TQI}.
During this kination period the inflaton is oblivious of the scalar potential
and, in this sense, out treatment is model-independent.

Throughout this paper we use natural units where
\mbox{$c=\hbar=1$} and Newton's gravitational constant is
\mbox{$8\pi G=m_P^{-2}$}, where \mbox{$m_P=2.4\times
10^{18}\,$GeV} is the reduced Planck mass.

\section{Basics of warm inflation}
In the warm inflationary models, the dissipative term appears as
an extra friction term in the evolution equation for the inflaton
field $\phi$, and as a source term for radiation $\rho_r$:
\begin{eqnarray}
\ddot \phi + (3 H + \Upsilon) \dot \phi + V' &=&0
\,,\label{eominf}\\
\dot \rho_r + 4 H \rho_r  &=& \Upsilon \dot \phi^2\,, \label{rad0}
\end{eqnarray}
where $V'$ denotes the derivative of the potential with respect to
the inflaton field.

Warm inflation is generically defined as the regime where
the temperature of the Universe is larger than the
Hawking temperature, $T > H$, since in this regime
the spectrum
of perturbations of the inflaton is generated from thermal
fluctuation, rather than from vacuum fluctuation as in the
cold inflation scenario.
Within this $T > H$ region there are two distinct
types of warm inflation regimes.
The first is when $\Upsilon>3H$
in which the dissipative effects result in extra
friction, making the field evolve slower. This is called
the \textit{strong dissipation} regime. The second is
when $\Upsilon\lesssim3H$, in which the
dissipative effects have a negligible influence on the evolution of the
inflaton, and so it then evolves as in the cold inflationary
scenario.  However since $T > H$, the fluctuations of
the inflaton are still thermal.
This is called the \textit{weak dissipation} regime.

With the dissipation coefficient $\Upsilon$ taken into account,
the slow roll equations of motion for $\phi$ and $\rho_r$ are
\begin{equation}\label{phidot}
  \dot\phi\simeq\frac{-V'}{(3H+\Upsilon)}\,,
\end{equation}
\begin{equation}\label{rad}
\rho_r\simeq\frac{\Upsilon}{4H}\,\dot{\phi}^2\simeq\frac{\Upsilon|V'|^2}
{4H(3H+\Upsilon)^2}\,,
\end{equation}
where the last equation holds provided the source term
$\Upsilon\dot{\phi}^2$ in Eq.~(\ref{rad0}) dominates. It then
follows that the produced radiation density is determined by the
kinetic density of the field.

The problem is understanding how the inflaton loses energy during
inflation from first principles in quantum field theory.
Considerable work has been done to address this problem for the
case where the entire system remains close to thermal equilibrium
\cite{Berera:1998gx,br,Moss:2006gt} Recently, this equilibrium
approach was developed for the low-temperature regime
\cite{Moss:2006gt}, which will prove useful for the analysis in
this paper. In this, the dissipation coefficient $\Upsilon$ is
computed in a class of supersymmetric models. The dissipation
mechanism is based on a two-stage process \cite{br}. The inflaton
field couples to heavy bosonic fields, $\chi$ and fermionic fields
$\psi_{\chi}$,  which then decay to light degrees of freedom.
These light degrees of freedom thermalize to become radiation. The
simplest superpotential containing such an interaction structure
is
\begin{equation}
W=
g\Phi X^2+hXY^2,
\label{W}
\end{equation}
where $\Phi$, $X$ and $Y$ denote superfields, and $\phi$, $\chi$
and $y$ refer to their bosonic components. Such an interaction
structure is common in many particle physics SUSY models during
inflation, the field $y$ and its fermionic partner $\bar{y}$
remain massless, whereas the field $\chi$ and its fermion partner
$\psi_{\chi}$ obtain their masses through their couplings to
$\phi$, namely \mbox{$m_{\psi_{\chi}}=m_\chi=g\phi$}. The regime
of interest is when \mbox{$m_\chi,m_{\psi_{\chi}} > T > H$}, and
this defines what is referred to here as the low-temperature
regime. For this regime the dissipation coefficient, when the
superfields $X$ and $Y$ are singlets, is found to be
\cite{Moss:2006gt}
\[\label{ups}
\Upsilon\simeq0.64\,g^2h^4\left(\frac{g\phi}{m_\chi}\right)^4
\frac{T^3}{m_\chi^2}\,,
\]
where $T$ is the temperature of the radiation bath, $\rho_r=C_rT^4$,
where $C_r=\pi^2 g_*/30$ and $g_*$ is the number of relativistic
degrees of freedom. In supersymmetric theories $C_r\simeq70$.
However, the superfields $X$ and $Y$ may belong to large
representations of a GUT group. In that case, the dissipation
coefficient picks up an extra factor ${\cal N}={\cal N}_\chi
{\cal N}^2_{\rm decay}$, where ${\cal N}_\chi$ is the multiplicity of the $X$
superfield and ${\cal N}_{\rm decay}$ is the number of decay channels
available in $X$'s decay.

Following this approach, it has been recently shown
\cite{BasteroGil:2006vr} that chaotic and hybrid inflation models
may support some $50$-$60$ $e$-foldings of warm inflation in the
strong dissipative regime. Using the slow roll parameters
\mbox{$\varepsilon\equiv-\dot{H}/H^2$} and $\eta\equiv m_P^2V''/V$, the
condition to have slow roll in warm inflation, for a potential
with $\varepsilon\ll \eta$, is found to be
\[\label{sr}
\frac{\eta}{1+\Upsilon/3H}<1\,.
\]
Hence, in the strong dissipative regime in which $\Upsilon/3H>1$,
the above holds even if $\eta>1$. As a result, the field drives
slow roll inflation even if it is not effectively massless, thus
evading the $\eta$-problem of inflation \cite{Berera:1999ws, Berera:2004vm}.

\section{A simple hilltop model}
In this paper we investigate a hilltop model
\cite{Alabidi:2005qi,hilltop1,hilltop2}. We consider the scalar
potential
\[\label{pot}
V=V_0-\frac{1}{2}|m^2|\phi^2+\ldots,
\]
where $V_0=3H^2m_P^2$, $m^2=V''(0)$, and the dots represent higher
order terms that become important only after relevant scales exit
the horizon during inflation. We assume that the field begins
close to the hilltop.
We also consider the dissipation
mechanism described above, and take $\Upsilon$ as given by
Eq.~(\ref{ups})
\[\label{ups2}
\Upsilon\simeq C_\phi\frac{T^3}{\phi^2}\,,
\]
where $C_\phi=0.64\,h^4{\cal N}$.
Writing $T=C_r^{-1/4}\rho_r^{1/4}$
and using Eq.~(\ref{rad}), the above equation becomes
\begin{equation}\label{Upsphi}
\Upsilon^{1/4}(3H+\Upsilon)^{3/2}\simeq C_\phi
C_r^{-3/4}\frac{|V'|^{3/2}}{(4H)^{3/4}\phi^2}\,,
\end{equation}
which determines $\Upsilon$ in terms of $\phi$ and the parameters
of the model. Working now backwards, we may use $\Upsilon$ and
Eq.~(\ref{rad}) to determine the radiation density and the
temperature. Then we work out the field dependence of the
ratios $T/H$, $m_{\chi}/T$, and $\Upsilon/3H$, which is summarised
in Table 1.
\begin{center}
\begin{tabular}{c|c|c|}
& Weak dissipation & Strong dissipation \\\hline
$\phantom{\Big{[}}\frac{T}{H}\propto$ & $|\eta|^2$ & $
|\eta|^{2/7}\phi^{4/7}$\\\hline
$\phantom{\Big[}\frac{m_\chi}{T}\propto$ & $|\eta|^{-2}\phi$ & $
|\eta|^{-2/7}\phi^{3/7}$\\\hline
$\phantom{\Big[}\frac{\Upsilon}{3H}\propto$ & $|\eta|^6\phi^{-2}$
& $ |\eta|^{6/7}\phi^{-2/7}$\\\hline
\end{tabular}\vspace{0.2cm}

\textbf{\footnotesize Table 1.}

\end{center}
It follows that if the field begins in the low-temperature limit,
its rolling away from the hilltop maintains the validity of the
low-temperature limit, because the ratios $T/H$ and $m_\chi/T$
grow as $\phi$ rolls down the potential. As a result, to describe
the entire evolution of the field during inflation it is only
necessary to chose appropriately the initial value of the field so
that the condition $H<T<m_\chi$ holds. Once this is satisfied, the
model allows a complete analytical study of the entire
inflationary stage. In Ref.~\cite{inprep} we discuss a string
inspired scenario in which the problem of the initial conditions
for warm inflation in the low-temperature limit may be motivated
from moduli trapping dynamics.

In this paper we find the parameter space where a warm
inflationary stage in the low-temperature limit, both in the weak
and strong dissipative regimes, results in enough inflation while
producing the observed spectrum of perturbations.

\section{The weak dissipative regime}
In our model, once the system evolves according to the weak
dissipative regime with $\Upsilon\lesssim3H$, it remains in such
limit for the rest of the evolution, as $\Upsilon$ decreases as
$\Upsilon\propto\phi^{-2}$ (see Table 1). Because in this case the
dissipative dynamics does not back-react on the evolution of the
inflaton field, the latter drives a substantial amount of
inflation only if the potential is sufficiently flat. We then
assume that the scalar potential satisfies $|\eta|\ll1$.

From Eq.~(\ref{rad}), we may compute the radiation density during
inflation
\[\label{radw}
\rho_r\simeq\frac{\Upsilon|V'|^2}{36H^3}\simeq4\times10^{-3}
\,C_\phi^4C_r^{-3}\,|\eta|^8H^4\,,
\]
which remains constant as long as $\eta$ is constant. The
temperature of the radiation is then
\[\label{temp}
T_{\rm inf}\simeq\frac1{4}\,C_\phi C_r^{-1}\,|\eta|^2H\,,
\]
which depends on the particle physics parameters $C_\phi$ and
$C_r$. However, to apply consistently the low-temperature regime
in the weak dissipation limit we need to impose the conditions
$T/H>1$, $m_\chi/T>1$ and $\Upsilon\lesssim3H$, which result in
the bounds
\[\label{lw1}
C_\phi>C_r|\eta|^{-2}\,,
\]
\[\label{lw2}
C_\phi<g\,C_r|\eta|^{-2}\frac{\phi}{H}\,,
\]
\[\label{lw3}
C_\phi\lesssim4\,C_r^{3/4}|\eta|^{-3/2}\left(\frac{\phi}{H}\right)^{1/2}\,,
\]
respectively. We consider that at least the `observable' amount of
inflation occurs in weak dissipation, and then take the field
value $\phi=\phi_*$ (when cosmological scales are exiting the
horizon) to satisfy the above constraints. Also, we require that
the thermal spectrum of perturbations reproduces the observed
spectrum. In the weak dissipation limit, the prediction for the
amplitude of the spectrum of curvature perturbations is
\cite{Moss:1985wn,Berera:1995wh}
\[
{\cal P_R}^{1/2}\simeq\frac{H}{|V'|}(3H+\Upsilon)\sqrt{TH}\,.
\]
Using Eq.~(\ref{temp}), the amplitude of the spectrum becomes
\[\label{spweak}
{\cal P_R}^{1/2}\sim C_\phi^{1/2} C_r^{-1/2}\,\frac{H}{\phi_*}\,.
\]

Owing to the constant temperature of the radiation when the
observable Universe leaves the horizon, the amplitude of the
spectrum is proportional to the one predicted from vacuum
fluctuations. As a result, the spectral index is the same as in
cold inflation for models with $\varepsilon\ll\eta$, namely
\[
n-1=\frac{d\ln{\cal P_R}}{d\ln k}\simeq-2|\eta|\,,
\]
which for negligible tensor perturbations requires
$|\eta|\simeq0.025$ to satisfy the observations
\cite{Spergel:2006hy,Alabidi:2005qi}. Inserting $\phi_*/H$ as
computed from Eq.~(\ref{spweak}) into Eqs.~(\ref{lw1}),
(\ref{lw2}) and (\ref{lw3}), we obtain the following bounds
\[\label{lw4}
C_\phi>C_r|\eta|^{-2}\,,
\]
\[\label{lw5}
C_\phi<g^2\,C_r|\eta|^{-4}{\cal P_R}^{-1}\,,
\]
\[\label{lw6}
C_\phi\lesssim6\,C_r^{2/3}|\eta|^{-2}{\cal P_R}^{-1/3}\,.
\]
From Eqs.~(\ref{lw4}) and (\ref{lw5}), the values of $g$
compatible with warm inflation are
\[
g>|\eta|{\cal P_R}^{1/2}\sim10^{-6}\,,
\]
which secures the existence of parameter space for almost any
physically interesting value of the coupling $g$. Also, for
$g>|\eta|{\cal P_R}^{1/3}\sim10^{-5}$, the bound in Eq.~(\ref{lw6})
is tighter than the one in Eq.~(\ref{lw5}). In order to keep the
discussion simple we neglect the latter.

After cosmological scales have exited the horizon, inflation
finishes when $\phi$ grows enough for higher order terms to become
important so that $|\eta|\sim1$. Following Ref.~\cite{hilltop2}, we take then
\mbox{$V=V_0-\frac1{2}|m^2|\phi^2+\delta V$}, where $\delta
V=-|V_n|\phi^n$ and $n>2$. Apart from terminating inflation, the
increase in $|\eta|$, growing from $|\eta|\simeq0.025$ to
$|\eta|\sim1$, necessarily leads to an increase in the temperature
during inflation, as Eq.~(\ref{temp}) suggests. Using the field
value \mbox{$\phi_{_{\rm
end}}\sim(\frac{3H^2}{n(n-1)|V_n|})^{1/(n-2)}$} at $|\eta|\sim1$, we
compute the temperature at the end of inflation
\[\label{Tend}
T_{\rm end}\sim T_{\rm inf}\left(1+\frac{|\eta_0|^{-1}}{n-1}\right)^2\,,
\]
where \mbox{$\eta_0\equiv\eta(\phi=0)$}. We find that even though
higher order terms may raise the temperature substantially, the
system still evolves within the low-temperature limit.

We consider now that the correction $\delta V$ steepens the
potential enough for the field to become dominated by its kinetic
density after inflation, \mbox{$\rho\simeq\rho_{\rm
kin}\equiv\frac1{2}\dot{\phi}^2\propto a^{-6}$}. Such a phase is
called \textit{kination} \cite{kin}. The equation of motion of the
inflaton field during kination is
$\ddot{\phi}+(3H+\Upsilon)\dot{\phi}\simeq0$, and its evolution
then becomes oblivious to the potential. One can envisage that
this phase of kination would finish when $\Upsilon$ (being
initially small: $\Upsilon<3H$) dominated over $H$, since the
latter decreases drastically during kination, $H\propto a^{-3}$.
However, because $T\propto a^{-1}$ and $\phi$ continues growing
after inflation, from Eq.~(\ref{ups2}) it follows that $\Upsilon$
decreases even faster than $H$. Consequently, the phase of
kination lasts until the radiation bath (whose density decreases
as $\rho_r\propto a^{-4}$ after inflation) becomes comparable to
the kinetic density of the field. At that moment the Hot Big Bang
evolution is recovered\footnote{A similar scenario has been
recently discussed in the context of curvaton reheating
\cite{Bueno Sanchez:2007in}.}. The growth of the scale factor from
the end of inflation until reheating, $a_{\rm rh}$, is given by
\mbox{$(\rho_r)_{_{\rm end}}a_{\rm rh}^{-4}\sim V_0a_{\rm
rh}^{-6}$}, and the reheating temperature, \mbox{$T_{\rm
rh}=T_{\rm end}a_{\rm rh}^{-1}$}, is then
\[\label{reh}
T_{\rm rh}\sim C_r^{1/2}\frac{T^3_{\rm end}}{Hm_P}\,.
\]

In order not to violate the current bounds on gravitino
overproduction, this reheating temperature must be kept
sufficiently low. However, in the warm inflation scenario,
gravitinos are not only produced when the Universe is reheated.
Owing to the thermal bath during warm inflation, gravitinos may
also be produced at the end of inflation. The relevant temperature
then being given by Eq.~(\ref{temp}), rather than by
Eq.~(\ref{reh}). Nevertheless, in Ref.~\cite{BasteroGil:2004tg} it
is argued that the relevant temperature to constrain gravitino
overproduction is the temperature at reheating. The reason is that
the entropy produced by the decay of the inflaton during reheating
can dilute the gravitinos produced at the end of inflation
\cite{Scherrer:1984fd}. Under this assumption, the reheating
temperature must then be larger than the temperature at
Nucleosynthesis and less than current bound due to gravitino
overproduction \cite{Sarkar:1995dd}, set to
\mbox{$T_{\rm rh}\lesssim T_{m_{3/2}}\equiv 10^{7-9}\,{\rm GeV}$}
for a gravitino mass
\mbox{$m_{3/2}={\cal O}(1)\,{\rm TeV}$}. Using
Eqs.~(\ref{temp}), (\ref{Tend}) and (\ref{reh}), the condition
\mbox{$T_{\rm BBN}<T_{\rm rh}<T_{m_{3/2}}$} results in the bounds
\[\label{lw7.5}
C_\phi<4\left(
\frac{T_{m_{3/2}}m_P}{H^2}\right)^{1/3}\left(|\eta_0|+\frac1{n-1}
\right)^{-2}C_r^{5/6}\,,
\]
\[\label{lw8}
C_\phi>4\left(
\frac{T_{BBN}m_P}{H^2}\right)^{1/3}\left(|\eta_0|+\frac1{n-1}
\right)^{-2}C_r^{5/6}\,.
\]
We find that, even if the temperature relevant to constrain
gravitino overproduction is the temperature at the end of
inflation, $T_{\rm end}$, then there still exists substantial
parameter space available.

There is yet another bound to take into account, concerning the spectrum
of relic gravitational waves. Gravitational waves (GWs) do not couple to
the thermal background of warm inflation. Their spectrum is then
generated by quantum fluctuations. This spectrum, owing to the phase
of kination that follows after inflation, has a slope which grows
with the frequency for those modes re-entering the horizon during
kination \cite{GW}. High frequency gravitons crossing inside the
horizon during kination may then alter the light element abundances
predicted by Nucleosyntesis by increasing $H$. To avoid this we
require \cite{eta,GW}
\begin{equation}\label{gravBBN}
I\equiv h^2\int_{\rm k_{\rm BBN}}^{k_*}\Omega_{\rm GW}(k)d\ln
k\leq2\times10^{-6}\,,
\end{equation}
where $\Omega_{\rm GW}(k)$ is the GW density fraction with
physical momentum $k$ and $h=0.73$ is the Hubble constant $H_0$ in
units of $100$ km/sec/Mpc. Using the spectrum $\Omega_{\rm GW}(k)$
as computed in \cite{GW}, the above constraint can be written as
\cite{eta}
\begin{equation}\label{eq}
I\simeq h^2\alpha_{\rm
GW}\Omega_{\gamma}(k_0)\frac{1}{\pi^3}\frac{H^2}{m_P^2}
\left(\frac{H}{H_{\rm rh}}\right)^{2/3}\,,
\end{equation}
where $\alpha_{\rm GW}\simeq 0.1$ is the GW generation efficiency
during inflation, \mbox{$\Omega_{\gamma}(k_0)=2.6\times
10^{-5}h^{-2}$} is the density fraction of radiation at present on
horizon scales, and
\mbox{$H_{\rm rh}\sim(C_r^{1/2}T_{\rm rh}^2)/m_P$} is
the Hubble parameter at the time of reheating. Using
Eq.~(\ref{reh}), the bound in (\ref{gravBBN}) translates into
the bound
\[\label{CGW}
C_\phi>\,C_r^{3/4}\,|\eta|^{-2}\,,
\]
which is roughly as strong as the bound in Eq.~(\ref{lw4}).

\begin{figure}[htbp]
\hspace{2.75cm} \epsfig{file=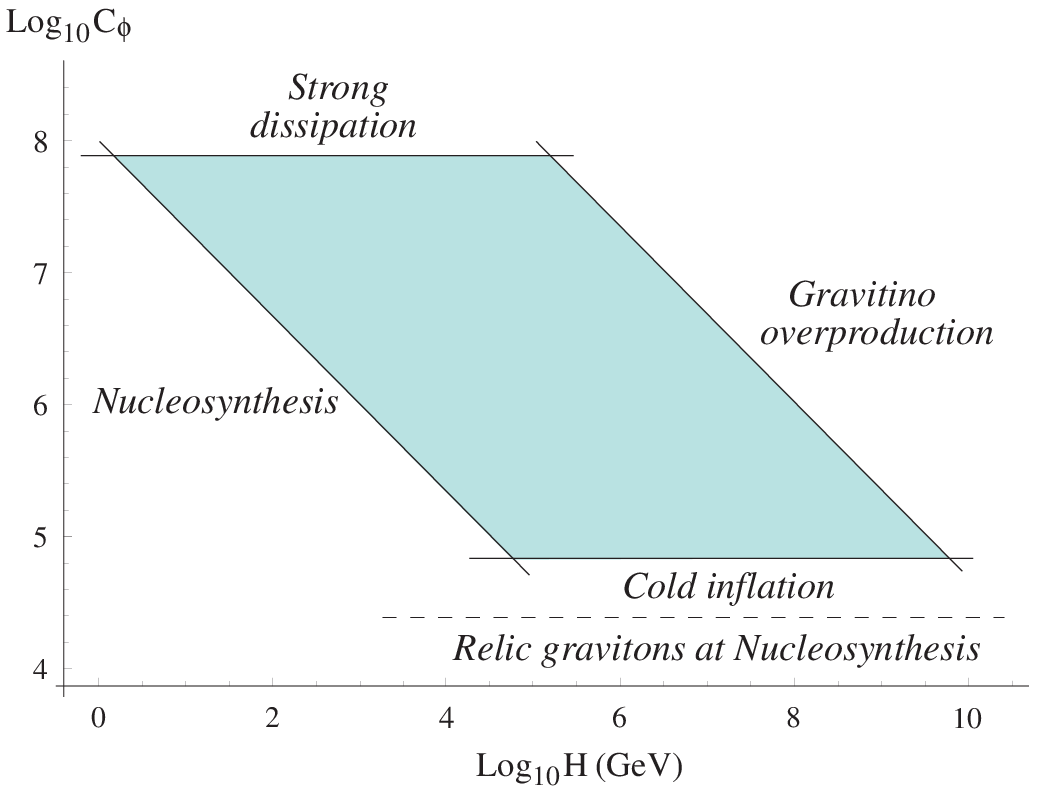,width=8.5cm} \caption{$\log
C_\phi$-$\log H$ plane in the weak dissipation limit. After the
relevant scales leave the horizon, inflation is terminated by
corrections of the potential \mbox{$\delta V=-|V_n|\phi^n$}. The
case shown correspond to a correction term with $n=4$. The
available parameter space (shaded area) is constrained by
gravitino
overproduction (using \mbox{$T_{m_{3/2}}\equiv10^{7-9}{\rm GeV}$}%
), the Nucleosynthesis
bound (using \mbox{$T_{\rm BBN}\sim{\rm MeV}$}), and the relic
graviton abundance at Nucleosynthesis, Eqs.~(\ref{lw7.5}),
(\ref{lw8}) and (\ref{CGW}), respectively. Consistency with warm
inflation imposes a lower bound on the amount of dissipation,
whereas the strong dissipation regime results in an upper bound,
Eqs.~(\ref{lw4}) and (\ref{lw6}) respectively.}\label{fig1}
\end{figure}

All bounds considered, the available parameter space leading to a
successful inflationary cosmology is depicted in Fig.~\ref{fig1}.
The parameter $C_r$ counting the number of relativistic degrees of
freedom has only a relative impact on the results. The only
significant effect being just a slight displacement of the allowed
region downwards \{upwards\} for lower \{higher\} values of the parameter.


%

We may summarize the results saying that even if the dissipation
mechanism does not bring the system into the strong dissipative
regime, there is still enough dissipation to maintain a radiation
bath $\rho_r$ which may reheat the Universe after inflation to
temperatures high enough to satisfy the Nucleosynthesis bound, but
not so high as to challenge the current bounds on gravitino
overproduction. There is a great deal of parameter space for
$C_\phi$, which also lies in the range of interest to GUT
theories. Also, owing to the phase of kination that follows after
inflation, the Hubble scale may be lowered down to the ${\rm GeV}$
scale \cite{Bueno Sanchez:2007in}.
Decreasing $H$ during inflation even further would move the
system into the strong dissipative regime, as $\Upsilon$ would
become important when compared to $3H$.

\section{The strong dissipative regime}
We consider now the case in which $\Upsilon$ is large enough for
the system to remain in strong dissipation until the end of
inflation. In this case, Eq.~(\ref{rad}) implies that
\mbox{$\rho_r>\rho_{\rm kin}$} during the inflationary stage. Therefore,
inflation finishes when the radiation density has grown enough for
the Universe to become radiation dominated, i.e. when
$\rho_r(\phi_{\rm end})\sim H^2m_P^2$. In this case, it is not
necessary to
invoke higher order terms in $V(\phi)$ to terminate inflation, and
we consider that $|\eta|$ remains constant until inflation is
finished.

Using Eqs.~(\ref{rad}) and (\ref{Upsphi}), we compute the
radiation density in strong dissipation, whose temperature is
\[
T\simeq (C_\phi
C_r)^{-1/7}\,|\eta|^{2/7}\left(\frac{\phi}{H}\right)^{4/7}H \,.
\]
Now we may compute $\phi_{\rm end}$ using the condition
\mbox{$\rho_r(\phi_{\rm end})\sim H^2m_P^2$},
\[
\phi_{\rm end}\sim
C_r^{-3/16}C_\phi^{1/4}\,|\eta|^{-1/2}\left(\frac{H}{m_P}\right)^{1/8}m_P\,.
\]
Inserting this into $\Upsilon(\phi_{\rm end})> 3H$ to have strong
dissipation until the end of inflation we obtain
\begin{equation}\label{radom}
C_\phi>4\, C_r^{3/4}|\eta|^{-2}\left(\frac{m_P}{H}\right)^{1/2}\,.
\end{equation}

To apply this scenario consistently we need to impose the
low-temperature conditions. The analogous of Eqs.~(\ref{lw1}),
(\ref{lw2}), and (\ref{lw3}) are now
\begin{equation}\label{TappH}
C_\phi<C_r^{-1}\,|\eta|^2\left(\frac{\phi}{H}\right)^4 \,,
\end{equation}
\begin{equation}\label{L}
C_\phi>g^{-7}C_r^{-1}\,|\eta|^2\left(\frac{\phi}
{H}\right)^{-3}\,,
\end{equation}
\[\label{upsH}
C_\phi>C_r^{3/4}\,|\eta|^{-3/2}\left(\frac{\phi}{H}
\right)^{1/2}\,.
\]

Exactly as we did in the weak dissipation regime, we take $\phi$ in
the above constraints when the observable Universe is leaving the
horizon, at $\phi=\phi_*$, and then fix $\phi_*$ by using the
amplitude of the spectrum of curvature perturbations. In the case
of strong dissipation the amplitude of the spectrum is given by
\cite{Berera:1999ws}
\[
{\cal P_R}^{1/2}\simeq\frac{H^2}{|V'|}\left(\frac{\pi
\Upsilon}{12H}\right)^{1/4}(3H+\Upsilon)\sqrt{\frac{T}{H}}\,,
\]
which in our model results in
\begin{equation}\label{pert}
{\cal P_R}^{1/2}\simeq0.4\,C_\phi^{9/14}C_r^{-17/28}
|\eta|^{3/14}\left(\frac{H}{\phi_*}\right)^ {15/14}\,.
\end{equation}
Using this to compute $(\phi_*/H)$, the bounds in
Eqs.~(\ref{TappH}), (\ref{L}), and (\ref{upsH}) taken at
$\phi=\phi_*$ become
\[\label{lw9}
C_\phi>C_r^{7/3}\,|\eta|^{-2}{\cal P_R}^{4/3}\,,
\]
\[\label{lw10}
C_\phi> g^{-5/2}C_r^{1/4}|\eta|^{1/2}{\cal P_R}^{1/2}\,,
\]
\[\label{lw11}
C_\phi>C_r^{2/3}|\eta|^{-2}{\cal P_R}^{-1/3}\,.
\]

In the range of $C_r$ of interest to particle physics models, the
bound in Eq.~(\ref{lw9}) is weaker than the one in Eq.~(\ref{lw11}).
Also, the bound in Eq.~(\ref{lw10}) is tighter than the one in
Eq.~(\ref{lw11}) for $|\eta|>g{\cal P_R}^{-1/3}$. Note that in the
strong dissipation regime it is possible to have slow roll inflation
with $|\eta|>1$.

After the relevant cosmological scales exit the horizon, some
$N_*\simeq50$ $e$-foldings must follow. Using Eq.~(\ref{phidot}),
we obtain
\begin{equation}\label{nw}
N_*\simeq 1.6\,C_\phi^{4/7}C_r^{-3/7}|\eta|^{-1/7}\left[
\left(\frac{H}{\phi}\right)^{2/7}\right]_{\phi_{\rm end}} ^{\phi_*}\,,
\end{equation}
which depends on $H$. Therefore, $H$ must be tuned
according to this equation. We obtain
\[\label{H}
\left(\frac{H}{m_P}\right)^{1/4}\sim
1.2\,\frac{C_r^{13/120}{\cal P_R}^{2/15}}{C_\phi^{1/10}|\eta|^{1/5}}-
0.6\frac{C_r^{3/8}N_*}{C_\phi^{1/2}}\,.
\]

\begin{figure}[htbp]
\hspace{5cm}\epsfig{file=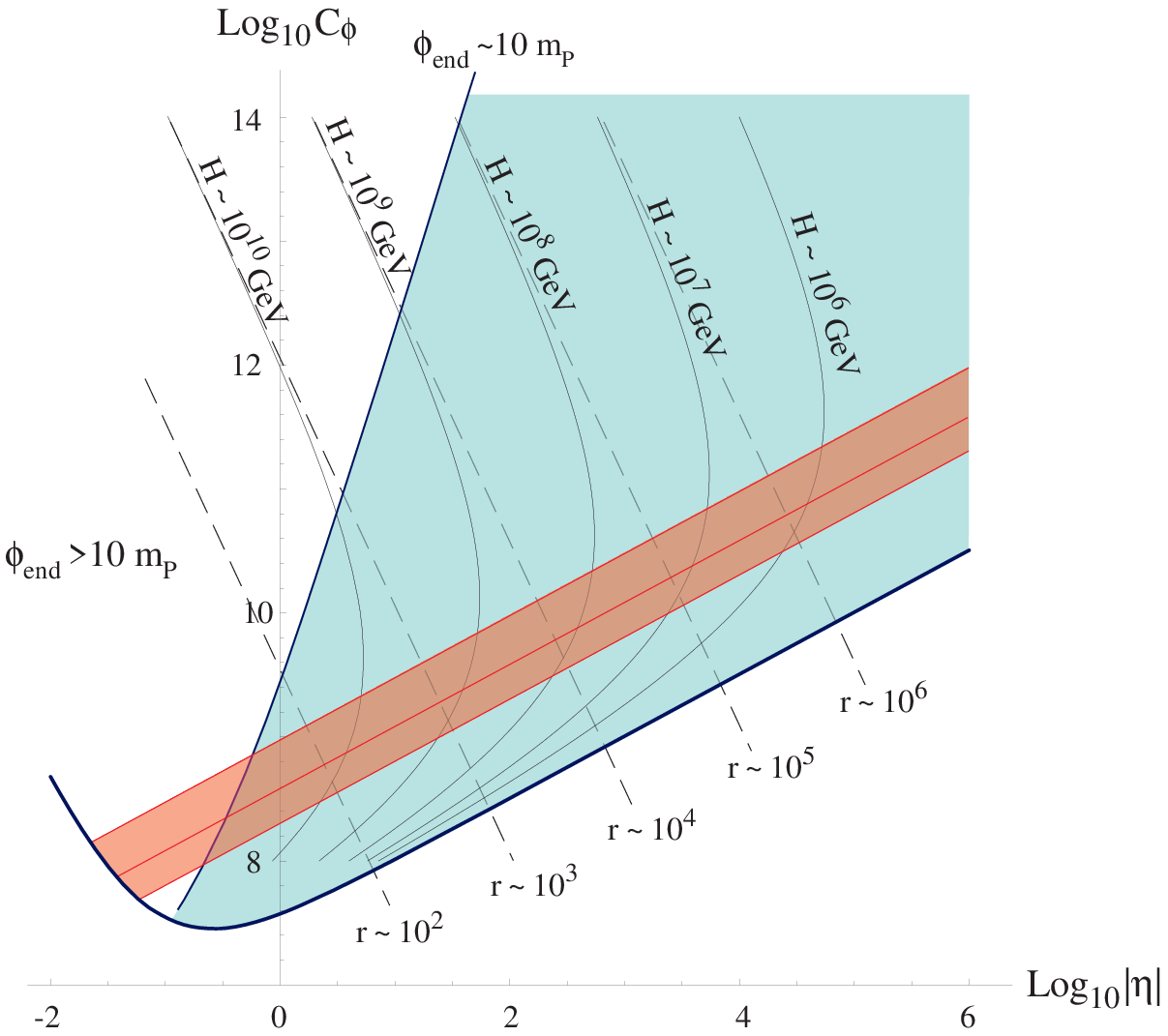, height=7.5cm}\caption{The
figure shows the $\log C_\phi$-$\log |\eta|$ plane. The light grey
(light green) area represents the space where the thermal spectrum
of perturbations matches the observed amplitude $N_*\sim50$
$e$-foldings before the end of inflation, with \mbox{$\phi_{\rm
end}<10\,m_P$}. Within the dark grey (reddish) band, the resulting
value of the spectral index is within the 1-$\sigma$ window:
\mbox{$n=0.948^{+0.016}_{-0.015}$}, as inferred by WMAP+SDSS data
in the $\Lambda$CDM model for negligible tensor
perturbations.
In the graph lines of constant $H$ and
\mbox{$r\equiv\Upsilon/3H$} are also depicted.}\label{fig2}
\end{figure}

If we neglect the negative contribution to $H$, the bound in
Eq.~(\ref{radom}) (coming from $\Upsilon(\phi_{\rm end})>3H$, which
ensures radiation domination after inflation) is equivalent to the one
in Eq.~(\ref{lw11}) (corresponding to $\Upsilon(\phi_*)>3H$).
This is expected, because owing to the
extra friction provided by the dissipation term, the inflaton
field varies very little during the last $N_*\simeq50$
$e$-foldings. Hence the ratio
$\Upsilon/3H\propto\phi^{-2/7}$ does not change substantially. We
may also use the above equation to eliminate $H$ from
Eq.~(\ref{radom}), obtaining
\begin{equation}\label{reg}
C_\phi^{2/5}> C_r^{4/15}{\cal P_R}^{-2/15}\left(1.6
|\eta|^{-4/5}+\frac{N_*|\eta|^{1/5}}{2}\right)\,.
\end{equation}
This bound determines the region in the $\log C_\phi$-$\log|\eta|$
plane where inflation results in a spectrum of superhorizon
perturbations with the observed amplitude. Once this bound is
satisfied, so is that in Eq.~(\ref{lw11}). However, the space
determined by the above must be further constrained by the bound
in Eq.~(\ref{lw10}) if
\mbox{$g<C_r^{-1/6}{\cal P_R}^{1/3}2N_*^{-1}\sim10^{-5}$}. Excluding
this case (i.e. for \mbox{$g\gtrsim 10^{-5}$}), the available
parameter space is completely determined by Eq.~(\ref{reg}).
This parameter space is depicted in Fig.~\ref{fig2}, where
we also enforced the requirement that
\mbox{$\phi_{\rm end}\lesssim m_P$} because over
super-Planckian distances in field space, the scalar potential
is not well understood.

The spectral index $n-1=\frac{d\ln{\cal P_R}}{d\ln k}$, upon
using Eq.~(\ref{pert}), becomes
\begin{equation}
n\simeq1-\frac{7\,C_r^{4/15}|\eta|^{1/5}}{2\,C_\phi^{2/5}{\cal P_R}^{2/15}}
\,.
\end{equation}
The dark grey (reddish) strip in Fig.~\ref{fig2} encompasses
the 1-$\sigma$ result
for the spectral index, \makebox{$n=0.948^{+0.016}_{-0.015}$},
obtained from WMAP+SDSS data in the $\Lambda$CDM model for
$\varepsilon\ll1$ \cite{Spergel:2006hy}. The running of $n$ is
\begin{equation}
  \alpha=\frac{dn}{d\ln k}\simeq -0.13 (n-1)^2\,,
\end{equation}
which is of order $10^{-4}$ within the 1-$\sigma$ window.

Non-Gaussian effects during warm inflation have been
studied \cite{Gupta:2002kn,Moss:2007cv} and it is interesting to see
what these analysis predict for this model.
In the strong dissipative regime it was shown in \cite{Moss:2007cv}
that entropy fluctuations during warm inflation play
an important role in generating non-Gaussianity, with
the prediction
\begin{equation}
  -15 \ln \left(1+ \frac{r}{14}\right) - \frac{5}{2} \alt f_{\rm NL} \alt \frac{33}{2} \ln \left(1+ \frac{r}{14}\right) - \frac{5}{2} \,,
\end{equation}
where $f_{\rm NL}$ is the non-linearity parameter and $r \equiv
\Upsilon/3H$. For the warm inflation results in Fig. (\ref{fig2}),
$r$ ranges from $10$ to $10^6$, and this  implies from the above
equation that $|f_{\rm NL}|$ ranges from $10$ to $180$.  This is
an interesting result in light of the recent WMAP analysis of the
third year CMB data \cite{Yadav:2007yy} which find $26.9 < f_{\rm
NL} < 146.7$ at $95\%$ confidence level. This corresponds to a
rejection of a Gaussian spectrum ($f_{\rm NL}=0$) at the 99.5\%
confidence level. If these first WMAP results are confirmed by
future data, and in particular by data from Planck surveyor
satellite \cite{planck}, this would exclude conventional cold
inflation models which generally yield very low values of $f_{\rm
NL} \stackrel{<}{\sim} 1$. On the other hand, the strong
dissipative warm inflation regime, such as the one found in this
hilltop model, would be consistent with this recent WMAP analysis.
For example in this case the WMAP upper limit on the parameter
$f_{\rm NL} \alt 150$ translates in an upper limit on $r \alt
1.4\times 10^5$, and therefore from Fig.~\ref{fig2} on a lower
limit on the scale of inflation given by \mbox{$H \gtrsim 10^7$
GeV}.

Summarizing, it is possible to match the amplitude of the spectrum
of perturbations in a large region of the $\log C_\phi$-$\log
|\eta|$ plane. Remarkably, the observations may be matched for
$|\eta|\gtrsim1$, thus avoiding the $\eta$-problem of inflation.
The number of fields required in this case  (taking $h^2\sim4\pi$
in $C_\phi=0.64h^4{\cal N}$) is large, ${\cal N}\sim10^{6-7}$. The
scale of inflation needed in the model to match observations
requires $H\gtrsim 10^6{\rm GeV}$, which challenges present bounds
on gravitino overproduction, however see Ref.~\cite{Hall:2007qw}.
Reducing further the Hubble scale during inflation is only
possible by using $|\eta|\gg1$, which requires $C_\phi\gg10^8$.
A low Hubble scale results in a larger value of the
non-Gaussian parameter $f_{\rm NL}$, which approaches the observational
bounds and may well be observable in the near future \cite{planck}.

\section{Conclusions}

We have studied warm inflation in the context of hilltop models.
There are important differences between this type of warm
inflation and the existing literature. This is because the value
of the inflaton field increases during inflation as it rolls down
the potential hill, which results in a decreasing dissipation
coefficient $\Upsilon$. This, ensures that the validity of the
low-temperature approximation persists throughout the evolution of
the system, but also tightens the constraints on the amount of
dissipation needed for the field to drive enough inflation
\cite{BasteroGil:2006vr}. During inflation the dissipation of the
inflaton's energy generates a thermal bath of roughly constant
temperature $T$, larger than the Hawking temperature. Hence, the
curvature perturbation is due to thermal instead of quantum
fluctuations. The Universe is reheated when this thermal bath
dominates.

When \mbox{$\Upsilon\lesssim 3H$} warm inflation occurs in the
so-called weak dissipative regime, when dissipation does not
affect the dynamics of the inflaton. In this case inflation has to
be terminated by higher-order terms in the scalar potential, which
steepen its slope. We have studied this case assuming that a brief
period of kination follows inflation, until reheating when the
radiation bath takes over the kinetic density of the inflaton.
This is reasonable to expect in case the potential is of runaway
type, as is the case for string moduli fields. Such a potential
has been considered in models of quintessential inflation
\cite{eta,TQI}. We have found that there is ample parameter space
for warm hilltop inflation in the weak dissipative regime (see
Fig.~\ref{fig1}) which interpolates between the usual cold
inflation case and the strong dissipative regime. The parameter
space allows low-scale inflation with $H$ as low as 1~GeV, in
accordance to Ref.~\cite{Bueno Sanchez:2007in}. This also implies
that the reheating temperature is low enough not to result in
gravitino overproduction. In this regime
\mbox{$T/H\simeq\,$constant}, which means that the curvature
perturbation depends only on $H$ as in cold inflation.
Consequently, the spectral index is identical to the cold hilltop
inflation case: \mbox{$n\approx 1-2|\eta|$}, where $|\eta|\simeq
0.025$ in order to account for the CMB observations. Hence, a mild
tuning of the curvature of the potential at the top of the hill is
required, corresponding to the usual $\eta$-problem of inflation.

When \mbox{$\Upsilon>3H$} warm inflation occurs in the so-called
strong dissipative regime, when the dissipation does control the
dynamics of the inflaton field. In this case, the temperature of
the thermal bath follows the growth of the inflaton as
\mbox{$T\propto\phi^{4/7}$}. Thus inflation can end with prompt
reheating, when the radiation density overtakes the potential
density of the field. Kination or higher order terms in the
potential need not be considered. The extra friction in the
variation of the inflaton due to the dominant dissipation term
allows slow-roll inflation with \mbox{$|\eta|\sim 1$} overcoming
thereby the $\eta$-problem of inflation. We have found that,
despite such a large value of $|\eta|$ there is considerable
parameter space where the spectral index of the curvature
perturbation agrees with CMB observations (see Fig.~\ref{fig2}).
However, inflation needs to take place at high energies, which
challenges gravitino overproduction constraints, unless
\mbox{$C_\phi\sim 10^2{\cal N}$} is rather large. Because $\cal N$
is determined by the field content of the theory this scenario is
best realised in the context of string theories with a large
number of degrees of freedom.
In addition, strong dissipation
during inflation can
result in substantial non-Gaussianity in the perturbation
\cite{Moss:2007cv}, which consistent with the recent WMAP analysis
on non-Gaussianity \cite{Yadav:2007yy} and
may become detectable in the near future by the Planck mission
\cite{planck}.
However, the upper limit on the amount of non-Gaussianity imposes
a lower bound on the Hubble scale, excluding the possibility
of low-scale inflation.

The above show that warm hilltop inflation is indeed possible both
with strong and weak dissipation. The scenario is distinct
compared to both cold hilltop inflation and also to the other
types of warm inflation studied in the literature
\cite{BasteroGil:2006vr,Taylor:2000ze,Berera:1998px}.
With weak dissipation low-scale inflation
is possible, while strong dissipation overcomes the
$\eta$-problem. In both cases, the Universe is reheated by the
thermal bath due to dissipation. This offers the intriguing
possibility that the inflaton does not decay after inflation, in
which case it may survive
until late times and play the role of quintessence. 

\begin{thebiblio
}{99}

\bibitem{book}
A.R.~Liddle and D.H.~Lyth,
{\em Cosmological Inflation and Large Scale Structure},
(Cambridge University Press, Cambridge U.K., 2000).

\bibitem{mssm}
R.~Allahverdi, K.~Enqvist, A.~Jokinen and A.~Mazumdar,
  JCAP {\bf 0610} (2006) 007;
J.~C.~Bueno Sanchez, K.~Dimopoulos and D.~H.~Lyth,
  JCAP {\bf 0701} (2007) 015.

\bibitem{Spergel:2006hy}
  D.~N.~Spergel {\it et al.}  [WMAP Collaboration],
  Astrophys.\ J.\ Suppl.\  {\bf 170} (2007) 377.

\bibitem{hybrid}
A.~D.~Linde,
  Phys.\ Rev.\  D {\bf 49} (1994) 748.

\bibitem{Copeland:1994vg}
E.~J.~Copeland, A.~R.~Liddle, D.~H.~Lyth, E.~D.~Stewart and D.~Wands,
Phys.\ Rev.\ D {\bf 49}, 6410 (1994).

\bibitem{shybrid}
G.~Lazarides, R.~K.~Schaefer, and Q.~Shafi,
Phys.\ Rev.\ D {\bf 56}, 1324 (1997).

\bibitem{Alabidi:2005qi}
  L.~Alabidi and D.~H.~Lyth,
  JCAP {\bf 0605} (2006) 016.

\bibitem{new}
  A.~D.~Linde,
  Phys.\ Lett.\  B {\bf 108} (1982) 389.

\bibitem{hilltop1}
  L.~Boubekeur and D.~H.~Lyth,
  JCAP {\bf 0507} (2005) 010.

\bibitem{hilltop2}
  K.~Kohri, C.~M.~Lin and D.~H.~Lyth,
  JCAP {\bf 0712} (2007) 004.

\bibitem{Dine:1995uk}
  M.~Dine, L.~Randall and S.~D.~Thomas,
  Phys.\ Rev.\ Lett.\  {\bf 75} (1995) 398;
  Nucl.\ Phys.\  B {\bf 458} (1996) 291.

\bibitem{FR}
  A.~Linde,
  JHEP {\bf 0111} (2001) 052.

\bibitem{curv}
  D.~H.~Lyth and D.~Wands,
  Phys.\ Lett.\  B {\bf 524} (2002) 5;
K.~Dimopoulos and D.~H.~Lyth,
  Phys.\ Rev.\  D {\bf 69} (2004) 123509.

\bibitem{trap}
  L.~Kofman, A.~Linde, X.~Liu, A.~Maloney, L.~McAllister and E.~Silverstein,
  JHEP {\bf 0405} (2004) 030.

\bibitem{warm}
  A.~Berera,
  Phys.\ Rev.\ Lett.\  {\bf 75} (1995) 3218;
Phys. Rev. D{\bf 54} (1996) 2519;
Phys.\ Rev.\  D{\bf 55} (1997) 3346.

\bibitem{Berera:1999ws}
A.~Berera,
Nucl.\ Phys.\ B {\bf 585}, 666 (2000).

\bibitem{Berera:2004vm}
  A.~Berera,
hep-ph/0401139.

\bibitem{Gaillard:1995az}
M.~K.~Gaillard, H.~Murayama and K.~A.~Olive,
Phys.\ Lett.\ B {\bf 355}, 71 (1995);
%
C.~F.~Kolda and J.~March-Russell,
Phys.\ Rev.\ D {\bf 60}, 023504 (1999).

\bibitem{randall}
N.~Arkani-Hamed, H.~C.~Cheng, P.~Creminelli and L.~Randall,
JCAP {\bf 0307}, 003 (2003).

\bibitem{ci} A. Linde, Phys. Lett. {\bf 129B}, 177 (1983).

\bibitem{kin}
B.~Spokoiny,
Phys.\ Lett.\ B {\bf 315} (1993) 40;
M.~Joyce and T.~Prokopec,
Phys.\ Rev.\ D {\bf 57} (1998) 6022;
D.~J.~H.~Chung, L.~L.~Everett and K.~T.~Matchev,
0704.3285 [hep-ph]; E.~J.~Chun and S.~Scopel,
0707.1544 [astro-ph].

\bibitem{eta}
K.~Dimopoulos,
Phys.\ Rev.\ D {\bf 68} (2003) 123506.

\bibitem{TQI}
  J.~C.~Bueno Sanchez and K.~Dimopoulos,
  JCAP {\bf 0710} (2007) 002;
  Phys.\ Lett.\  B {\bf 642} (2006) 294
  [Erratum-ibid.\  B {\bf 647} (2007) 526].

\bibitem{Berera:1998gx}
  A.~Berera, M.~Gleiser and R.~O.~Ramos,
  Phys.\ Rev.\  D {\bf 58} (1998) 123508.

\bibitem{br} A. Berera and R. O. Ramos,
Phys. Rev. D{\bf 63}, 103509 (2001).

\bibitem{Moss:2006gt}
  I.~G.~Moss and C.~Xiong,
hep-ph/0603266.

\bibitem{BasteroGil:2006vr}
  M.~Bastero-Gil and A.~Berera,
  Phys.\ Rev.\  D {\bf 76} (2007) 043515.


\bibitem{inprep}
 J.~C.~Bueno Sanchez, M.~Bastero-Gil, A.~Berera and K.~Dimopoulos,
in preparation.

\bibitem{Moss:1985wn}
  I.~G.~Moss,
  Phys.\ Lett.\  B {\bf 154} (1985) 120.

\bibitem{Berera:1995wh}
  A.~Berera and L.~Z.~Fang,
  Phys.\ Rev.\ Lett.\  {\bf 74} (1995) 1912.

\bibitem{Bueno Sanchez:2007in}
 J.~C.~Bueno Sanchez and K.~Dimopoulos,
  JCAP {\bf 0711} (2007) 007.

\bibitem{GW}
M.~Giovannini,
Phys.\ Rev.\ D {\bf 60} (1999) 123511;
V.~Sahni, M.~Sami and T.~Souradeep,
Phys.\ Rev.\ D {\bf 65} (2002) 023518.

\bibitem{Sarkar:1995dd}
  For a review, see S.~Sarkar,
  Rept.\ Prog.\ Phys.\  {\bf 59} (1996) 1493;
%
  M.~Kawasaki, K.~Kohri and T.~Moroi,
  Phys.\ Lett.\  B {\bf 625} (2005) 7.

\smallskip

\bibitem{BasteroGil:2004tg}
  M.~Bastero-Gil and A.~Berera,
  Phys.\ Rev.\  D {\bf 71} (2005) 063515.

\bibitem{Scherrer:1984fd}
  R.~J.~Scherrer and M.~S.~Turner,
  Phys.\ Rev.\  D {\bf 31} (1985) 681.

\bibitem{Gupta:2002kn}
  S.~Gupta, A.~Berera, A.~F.~Heavens and S.~Matarrese,
  Phys.\ Rev.\  D {\bf 66} (2002) 043510.

\bibitem{Moss:2007cv}
  I.~G.~Moss and C.~Xiong,
  JCAP {\bf 0704} (2007) 007.

\bibitem{Yadav:2007yy}
A.~P.~S.~Yadav and B.~D.~Wandelt,
0712.1148 [astro-ph].

\bibitem{planck}  Planck Surveyor Mission:\\
  http://www.rssd.esa.int/Planck.

\bibitem{Hall:2007qw}
  L.~M.~H.~Hall and H.~V.~Peiris,
0709.2912 [astro-ph].

\bibitem{Taylor:2000ze}
  A.~N.~Taylor and A.~Berera,
  Phys.\ Rev.\  D {\bf 62} (2000) 083517.

\bibitem{Berera:1998px}
  A.~Berera, M.~Gleiser and R.~O.~Ramos,
  Phys.\ Rev.\ Lett.\  {\bf 83} (1999) 264;
%
  A.~Berera and T.~W.~Kephart,
  Phys.\ Rev.\ Lett.\  {\bf 83} (1999) 1084.

\end{thebiblio}
\end{document}